\documentclass[%
aps,%
prb,%
superscriptaddress,%
preprintnumbers,%
byrevtex%
]{revtex4}

\usepackage{ifthen}
\usepackage{color}
\usepackage{amssymb}
\usepackage{amsfonts}
\usepackage{amsmath}
\usepackage[dvips]{graphicx}

\def\ii{\textrm{i}\,\!}

\newcommand{\beann} {\begin{eqnarray*}}
\newcommand{\eeann} {\end{eqnarray*}}
\newcommand{\bea} {\begin{eqnarray}}
\newcommand{\eea} {\end{eqnarray}}
\newcommand{\labs} {\left\vert}

\newcommand{\lab} {\left\langle}
\newcommand{\rab} {\right\rangle}

\begin{document}
\date{October 10, 2003}

\title{Electron transport in carbon nanotube-metal systems: contact effects}

\author{N.~Ranjan}
\affiliation{Institut f{\"{u}}r Physikalische Chemie und Elektrochemie
Technische Universit{\"{a}}t Dresden, Germany }
\affiliation{Institute for Theoretical Physics, University of Regensburg,
D-93040 Regensburg, Germany}

\author{R.~Guti{\'e}rrez-Laliga}
\affiliation{Institute for Theoretical Physics, University of Regensburg,
D-93040 Regensburg, Germany}

\author{S.~Krompiewski}
\affiliation{Institute of Molecular Physics, Polish Academy of Sciences,
PL-60179 Pozna{\'n}, Poland}

\author{G.~Cuniberti}
\affiliation{Institute for Theoretical Physics, University of Regensburg,
D-93040 Regensburg, Germany}

\begin{abstract}
Carbon nanotubes (CNT) have a very large application potential in the rapid developing 
field of molecular electronics. 
Infinite single-wall metallic CNTs have theoretically a conductance of $4e^{2}/h$ because of 
the two electronic bands crossing the Fermi level.
 For finite size CNTs experiments have shown that other values are also 
possible, indicating a very strong influence of the contacts. 
We study electron transport in single- and double-wall CNTs 
contacted to metallic electrodes within the Landauer formalism combined with Green 
function techniques. We show that 
the symmetry of the contact region may lead to blocking of a transport channel. 
In the case of double-wall CNTs  with both inner and outer shells
 being metallic, non-diagonal self energy contributions from the electrodes may induce channel mixing,
 precluding a simple addition of the individual shell conductances.

\end{abstract}

\maketitle

\section{Introduction}
Carbon nanotubes belong to one of the most promising candidates in the era of modern nanoelectronics. 
They can be generated by wrapping a graphene sheet along different directions as given by the so called chiral vector \cite{saitoMG}. 
Interestingly, depending on the chiral vector, the tubes show markedly different electronic 
properties ranging from metallic to semiconducting. 
A considerable amount of theoretical and experimental research has been done to explore 
their varied interesting properties, which  range from very hard inert materials 
through good conductors to  storage devices\cite{Reich}.
Concerning the electronic transport properties of metallic single-wall 
carbon nanotubes (SWCNT) experiments\cite{Reich} have nicely demonstrated  ballistic transport 
 and conductance quantization with conductance values equal to 2$\times G_0$, $G_0=2e^2/h$ 
being the  conductance quantum  and the factor 2 in the first expression 
arising from two spin degenerated bands at the Fermi level.   
Similar quantization effects have been recently observed in multi-wall nanotubes (MWCNTs)
\cite{poncharalPZW,urbinaIAAJ}.
However, in contrast to the theoretical expected values conductance steps as low as 0.5$G_0$ or 1$\times G_0$ 
were found\cite{poncharalPZW}. Even under the usual assumption that only the outermost shell is the one contributing 
to transport, such small conductance values suggest that  transport channels may be partially or completely closed. 
Blocking of conductance channels 
in MWCNTs has been recently addressed in other theoretical works\cite{sanvitoYDC}. We will investigate 
in this paper conductance quantization in finite size CNTs contacted with metallic electrodes.  
 We will show an example of channel blocking and demonstrate that the total conductance of muti-wall CNTs 
 cannot be simply obtained by just adding the individual shell conductances.

\section{Theoretical model}
\begin{figure}[h]
\centerline{\includegraphics[width=.70\linewidth]{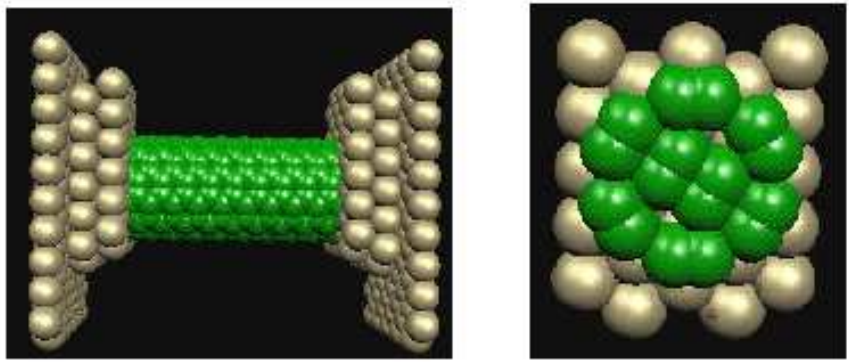}}
\caption{\label{fig:fig1} View of a (2,2)@(6,6)  double walled carbon nanotube 
sandwiched between two fcc(111) leads (upper panel)  and 
details of the contact region (lower panel). 
The first two layers of the electrodes belong to the scattering region while 
the third layer extends to infinity and is  part of the contact.
}
\end{figure}

We investigate electronic transport in a simple model system consisting of carbon nanotubes 
which are  attached to 
semi-infinite electrodes having an fcc(111)
geometry. 
A typical configuration is shown in Figure~\ref{fig:fig1}.

To describe the electronic structure of both subsystems we use a single-orbital nearest-neighbours 
tight-binding approach. It includes
the $\pi$-orbitals of the carbon atoms on the tube and s-like orbitals in the electrodes. The metal-CNT coupling
terms are set constant for all neighbours of a given carbon atom. The Hamiltonian is

\begin{eqnarray}
  H &=& H_{\cal M} + H_{\rm leads} + V_{{\rm leads},{\cal M}} \nonumber  \\ 
  H_{\cal M} &=&  {-t_{pp}\sum_{l,j} c^{\dagger}_{l}c_{j}}-\beta \sum_{\,l,\,j}{\cos\theta_{lj}}
e^{\frac{a-d_{\,l\,j}}{\delta}}c^{\dagger}_jc_l  \nonumber \\ 
  H_{\rm leads} &=& {\sum_{{ k}}\sum_{\alpha\in \mathrm{L,R}} \epsilon^{\alpha}_{k}d^{\dagger}_{{ k}\alpha}d_{{ k}\alpha}} \nonumber \\
  V_{{\rm leads},{\cal M}}& =& \sum_{i,{ k}}\sum_{ \alpha\in \mathrm{L,R}} V_{i,\alpha}c^{\dagger}_{i} d_{{ k}\alpha} +  \textrm{H.c.} \nonumber
 \end{eqnarray}
$H_{\cal M}$ is the CNT Hamiltonian. 
Its first term describes the intra-shell interaction with a hopping integral $t_{pp}$ 
which is set at the constant value of 2.66 eV. The second summand is the 
inter-shell interaction in the case of MWCNTs ($\beta=t_{pp}/k, k>1$). 
$\delta=0.45 {\AA} $ 
 is a normalizing factor 
, $a$ is the difference between the shell radii and $\theta_{ij}$ 
is the angle between the two $p_z$ orbitals on different shells. Finally, 
$H_{\rm leads}$ is the Hamiltonian of the electrodes and $ V_{{\rm leads},{\cal M}}$ is the mutual
interaction.

The linear conductance $G(E)$ can be related to the electronic transmission probability $T(E)$ 
 according to the Landauer formula (at zero temperature): $ {G} =G_0 T(E_{\rm F})$.
$T(E)$ can be calculated using Green function techniques\cite{datta} via
\begin{eqnarray}
T = {\rm Tr}_{\cal M} [ { G^{\dagger}_{{\cal M}}}{ \Gamma_R}{ G_{{\cal M}}}{ \Gamma_L} ]  \nonumber
\end{eqnarray}

$ G_{{\cal M}}$ is the Green function of the scattering region (in our case the CNT plus two surface layers) which can be calculated by means of the Dyson equation
\begin{eqnarray}
[ E{ 1_{{\cal M}}}-{ H_{\cal M}}-{ \Sigma_L}-{ \Sigma_R}]{ G_{\cal M}} ={ 1} \nonumber
\end{eqnarray}
The self-energies $\Sigma_{\alpha}= V^{\dagger}_{\alpha}g^{\alpha}V_{\alpha} ,\alpha=\mathrm{L,R} $ 
contain information 
on the electronic structure of the leads (via the surface Green function $g^{\alpha}$)
as well as on the electrode-scattering region coupling (via $V_{\alpha}$). 
Finally the spectral functions  $\Gamma_{\alpha}$ are related to the self-energies by 
$\ii \Gamma_{\alpha} = (\Sigma_{\alpha} - \Sigma_{\alpha}^{\dagger})$. We do not consider charge transfer 
effects between the tubes and the metallic electrodes.The use of  a single-orbital picture to describe the electrodes allows to write an analytic expression for 
the electrodes surface Green function in ${\bf k}$-space (assuming L=R)\cite{todorovGA}.
 \begin{eqnarray}
g({\bf k},E)&=&\frac{E-\epsilon({\bf k}) \pm \sqrt{(E-\epsilon({\bf k}))^2-4|V_{01}({\bf k})|^2 } } 
{2|V_{01}({\bf k})|^2 } \nonumber \\
\epsilon({\bf k})&=&2t_0 (\cos{k_x a}+2\cos{\frac{k_x a}{2}} \cos{ \frac{\sqrt{3}k_y a}{2} } ) \nonumber \\
 V_{01}({\bf k})&=& -t_{0} (2\cos({\frac{k_x a}{2}}) e^{{\frac{i k_{y} a} { 2\sqrt{3} }} }
+e^{{-\frac{ik_y a}{\sqrt{3}}}}) \nonumber 
\end{eqnarray}


\subsection{Electronic Transport in single- and double-wall nanotubes}
\begin{figure}[h]
\centerline{\includegraphics[width=.50\linewidth]{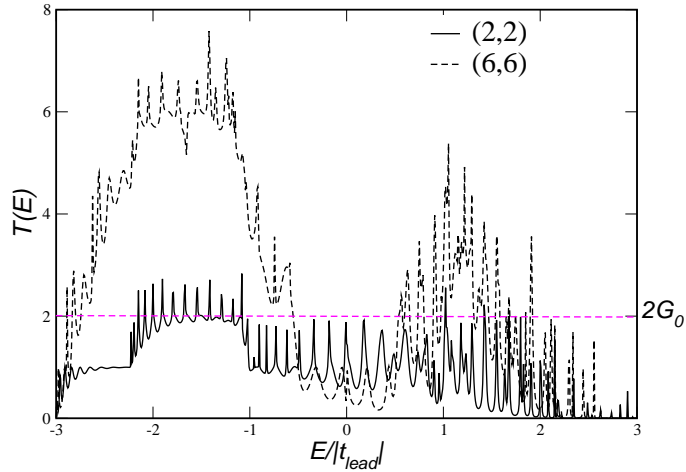}}
\caption{\label{fig:swcnt}
Electronic transmission of (2,2) and (6,6)  SWCNTs (10 unit cells). 
The strong oscillations are related to finite size effects. 
For (2,2) two transport channels  at the Fermi level are open, while 
for the (6,6) CNT only one channel does contribute.}
\end{figure}

We first consider the diameter dependence of the conductance for SWCNTs. In Fig. \ref{fig:swcnt} we show the 
transmission as a function of the energy for (2,2) and (6,6) finite size tubes (10 unit cells). 
As a reference we also plot the transmission of an infinite tube. 
For the latter clear quantization steps are obtained and the 
conductance around the Fermi level is 2$\times G_0$. The finite size tubes show however a more irregular, 
spiky  behaviour which can be related to finite size effects and to the 
lifting of some degeneracies as a result of the coupling to the electrodes. 
More importantly, while the (2,2) CNT shows conductance oscillations peaking at 2$\times G_0$, 
the (6,6) tube reaches only on average one quantum of conductance, i.e. a transport channel is 
apparently closed.

We can roughly understand what happens by representing  the electronic selfenergy into the 
eigenstate basis $\labs \Phi_{\sigma}\rab$ of an isolated CNT, with  
$\labs \Phi_{\sigma}\rab= \sum_{n\in CNT} {c_{n,\sigma} }\labs p_{z,n}\rab  $. 
After some manipulations ~\cite{gur01} 
and assuming a constant coupling of each carbon atom to its nearest neighbours metal atoms ~\cite{gur02,krompiewskiJMMM}, 
we arrive at the following expression:

\begin{eqnarray}
\Sigma_{\sigma\sigma'}(E)&=&|V|^2\sum_{{\bf k_{||} }} {g_{0,{\bf k_{||}}}(E)} \Lambda^{\dagger}_{\sigma}({\bf k_{||}})
\Lambda_{\sigma'}({\bf k_{||}})  \\
\Lambda_{\sigma}({\bf k_{||}})&=&
{{ \sum_{{\bf m_{||}{[\sl n]}}}} {\sum_{n}c_{n,\sigma}}{e^{i{\bf
k_{||}}{\bf m_{||}} }} } \nonumber
\end{eqnarray}

Notice that the index $n$ runs now over the CNT atomic slice in {\it direct} 
contact with the metal surface and the index 
${\bf m_{||}}[\sl n]$ denotes the nearest neighbours on the electrode surface of a given carbon atom $n$. 
The function  $\Lambda_{\sigma}({\bf k_{||}})$ contains information on the symmetries of the CNT wave functions 
via $c_{n,\sigma}$, and on the electrode surface topology, 
via the $e^{i{\bf k_{||}}{\bf m_{||}}} $ factor. 
We only need to look at the behaviour of  $\Lambda_{\sigma}({\bf k_{||}})$ for $\sigma=\pi,\pi^{*}$, since these 
are the two eigenstates crossing the Fermi point in metallic CNTs. 
Remembering that the expansion coefficients $c_{n,\sigma}$  of the
 $\pi$ and $\pi^{*}$ orbitals along the nanotube circumference comprising $2m$ atoms are proportional
 to $(+1)^{n}$ and $(-1)^n,\; \; n=1,\cdots ,2m$, respectively,
 the sums in  $\Lambda_{\sigma}({\bf k_{||}})$  can be performed.  As
a result we find that 
 $\Lambda_{\pi^{*}}({\bf k_{||}})$ identically vanishes for the (6,6) CNT while it 
is nonzero for the (2,2) tube. Hence the antibonding $\pi^{*}$ orbital does not couple to the electrodes for the 
(6,6) CNT and 
it thus gives no contribution to the 
conductance.

The next issue we have considered  is if  
the conductance of a DWCNT consisting of two armchair SWCNTs can be simply obtained by adding the 
corresponding conductances of the individual shells. If this holds then, accordning to our previous result, 
the conductance of a finite size
(2,2)$@$(6,6) DWCNT should yield 3$\times G_0$. 
Two factors can however modify this simple picture. 
One is the inter-shell coupling, the other is the CNT-electrode interface. 
We have just seen, that the latter can even induce channel blocking. 

\begin{figure}[h]
\centerline{\includegraphics[width=.50\linewidth,clip=true]{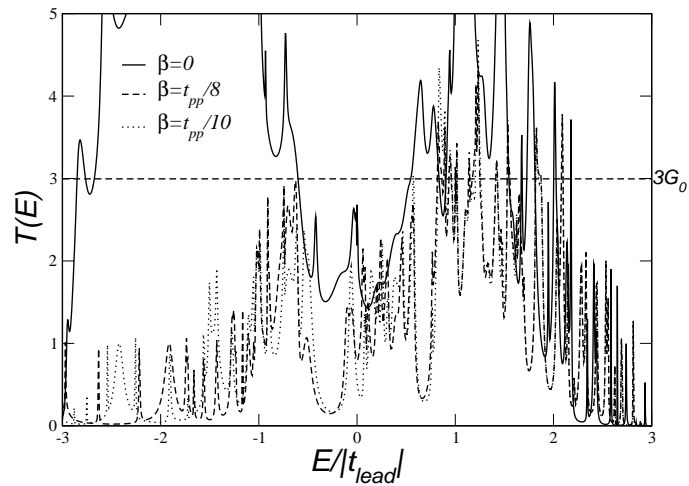}}
\caption{\label{fig:interwall} Energy dependent transmission of a 
 (2,2)@(6,6) DWCNT (10 unit cells) for different values of the inter-shell coupling $\beta$. The intra-shell 
hopping  $t_{pp} = 2.66eV$.}
\end{figure}

Figure~\ref{fig:interwall} shows the transmission function for different values of the inter-shell coupling parameter 
$\beta$. The main influence of $\beta$ is to introduce  mixing of the transport channels which is rather strong at 
energies far away from the Fermi level $E_{\rm F}$ 
and leads for some 
energies to a drastic reduction of the conductance when comparing with infinite tubes. 
The effect near the Fermi level is however less strong. 
Thus, for $\beta\neq 0$ 
the conductances can not simply be added since inter-shell interference effects must be considered. 
More interesting 
however is the behaviour for {\it zero} inter-shell interaction.
 Even in this case the total conductance near the Fermi level is not simply  3$\times G_0$ 
although it is larger than in the former case ($\beta\neq 0$).
The imperfect addition of conductances 
is now related to interference effects caused by non-diagonal contributions  of the electrodes 
self-energies, $\Sigma_{\sigma\neq\sigma^{'}}(E)$. As a result the transport channels are mixed in a similar way as for non-zero inter-shell 
coupling. Although there may be some special cases where conductances can be added, we can
 conclude that in general quantum interference effects induced by finite size effects (the existence of 
the metal-CNT interface) or by the coupling between the nanotube 
shells will preclude this simple view.

\section{Conclusions}
In this paper we have investigated quantum transport in finite size armchair single- and double-wall carbon nanotubes  contacted by metallic electrodes. We have shown that symmetries of the CNT-electrode coupling, hidden in 
the electronic self-energies may lead to suppression of transport channels, thus reducing the conductance around the Fermi level when comparing with the theoretical ideal case of infinite nanotubes. 
Moreover, for DWCNTs the simple approach of adding the single-shell 
conductances has been shown to be incorrect, even in the case of 
no inter-shell interactions. This can be traced back to interference effects induced by non-diagonal 
components of the self-energy. These results accentuate the important role played by the interface 
in determining electronic transport on  nanoscale systems.



\begin{thebibliography}{10}

\bibitem{saitoMG}
R. Saito, M.~S. Dresselhaus, and G. Dresselhaus, {\em Physical Properties of
  Carbon Nanotubes}, Imperial College Press  (1998).

\bibitem{Reich}
S. Reich, C. Thomsen, and J. Maultzsch, {\em Carbon Nanotubes: Basic Concepts
  and Physical Properties}, Wiley-VCH  (2004).

\bibitem{poncharalPZW}
S. Frank, P. Poncharal, Z.~L. Wang, and W.~A. de~Heer, {\em Carbon Nanotube
  Quantum Resistors}, Science {\bf 280},  1744  (1998).

\bibitem{urbinaIAAJ}
A. Urbina {\it et~al.}, {\em Quantum Conductance Steps in Solutions of
  Multiwalled Carbon Nanotubes}, Phys.\ Rev.\ Lett. {\bf 90},  106603  (2003).

\bibitem{sanvitoYDC}
S. Sanvito, Y.~K. Kwon, D. Tomanek, and C.~J. Lambert, {\em Fractional Quantum
  Conductance in Carbon Nanotubes}, Phys.\ Rev.\ Lett. {\bf 84},  1974  (2000).

\bibitem{datta}
S. Datta, {\em Electronic Transport in Mesoscopic Systems}, Cambridge
  University Press, Cambridge  (1995).

\bibitem{todorovGA}
T.~N. Todorov, G.~A.~D. Briggs, and A.~P. Sutton, {\em Elastic quantum
  transport through small structures}, J.\ Phys.-Condens.\ Matter {\bf 5},
  2389  (1993).

\bibitem{gur01}
, Using $\labs \Phi_{\sigma}\rab$ and the electrode wave functions $\labs {\bf
  k}\rab=\sum_{{\bf \ell_{||}}} {e^{i{\bf k_{||}{\bf \ell_{||}}}}} { f({\bf
  k_{\bot}})}\labs \chi_{{\bf \ell_{||}}}\rab$ we get:
  $\Sigma_{\sigma\sigma'}(E)=\sum_{{\bf k }} \lab
  \Phi_{\sigma}|V^{\dagger}|{\bf k}\rab g({\bf k},E)\lab{\bf
  k}|V|\Phi_{\sigma'}\rab$, which can be put in the form of Eq. (1) by defining
  ${G_{{\bf k_{||}}}(E)}=\sum_{{\bf k_{\bot}}} |f({\bf k_{\bot}})|^2 { g({\bf
  k_{||}},{\bf k_{\bot}} ,E)}$. $f({\bf k_{\bot}})$ is a function of the wave
  vector perpendicular to the metal surface, only, and the 2d vector ${\bf
  \ell_{||}}$ runs over the surface layer  .

\bibitem{gur02}
, This approximation can be justified because of the almost perfect fitting of
  the (2,2) and (6,6) CNTs lattice constants to the interatomic distances at
  the electrode surface ~\cite{krompiewskiJMMM}. Thus, each C atom interacts
  with just 3 atoms on the metal surface  .

\bibitem{krompiewskiJMMM}
S. Krompiewski, {\em Non-equilibrium transport in ferromagnetically contacted
  metallic carbon nanotubes}, Journal of Magnetism and Magnetic Materials {\bf
  In press},    (2004).

\end{thebibliography}

\end{document}